\documentclass[twocolumn.apjl]{emulateapj}
  \usepackage{amsmath}
  \usepackage{txfonts}
  \usepackage{latexsym}
  \usepackage{amsfonts}
  \usepackage{amssymb}
  \usepackage{amsbsy}
  \usepackage{natbib}
  \usepackage[english]{babel}
  \usepackage{color}

   \newcommand{\Rsun} {R_\odot}
   \newcommand{\fmax} {f_{\mathrm{max}}}

   \newcommand{\sR} {\textsf{R}}
   \newcommand{\sbR} {\textbf{\textsf{R}}}

   \newcommand{\vct}[1]  {\ensuremath{\boldsymbol{#1}}}    
   \renewcommand{\d} {\mathrm{d}}
   \newcommand{\Deriv} [2]    {\frac{\d   {#1}} {\d{#2}} }
   \newcommand{\Grad} {{\nabla}}
   \newcommand{\Div}  {{\nabla \cdot}}
   \newcommand{\vU} {\vct{U}}
   \newcommand{\vb} {\vct{b}}
   \newcommand{\vu} {\vct{u}}
   \newcommand{\vB} {\vct{B}}
   \newcommand{\rhat}  {\hat{\vct{r}}}
   \newcommand{\half}      {\frac{1}{2} }

\shorttitle{Turbulence-driven fast solar wind}
\shortauthors{Verdini et al.}

\begin{document}

\title{A turbulence-driven model for heating and acceleration of the
        fast wind in coronal holes}

\author{A.   Verdini,\altaffilmark{1}
        M.   Velli,\altaffilmark{2,3}
        W. H. Matthaeus,\altaffilmark{4}
        S.   Oughton,\altaffilmark{5}
        P.   Dmitruk,\altaffilmark{6}}
\altaffiltext{1}{Observatoire Royale de Belgique,
        3 Avenue Circulaire, 1180,
        Bruxelles, Belgium;
        e-mail: verdini@oma.be}
\altaffiltext{2}{Dipart. di Astronomia e Scienza dello Spazio,
        Univ.\  di Firenze,
        Largo E.\  Fermi 3, 50125, Firenze, Italy}
\altaffiltext{3}{Jet Propulsion Laboratory,
        California Institute of Technology,
        4800 Oak Grove Drive, Pasadena, CA 91109, USA}
\altaffiltext{4}
  	{Dept.\ of Physics and Astronomy,
	Univ.\ of Delaware, USA}
\altaffiltext{5}
  	{Dept.\ of Mathematics, Univ.\ of Waikato, Hamilton, New Zealand}
\altaffiltext{6}
 {Depart.\ F\'\i sica,
        Facultad de Ciencias Exactas y Naturales,
        Univ.\ de Buenos Aires-Conicet, Argentina}

\begin{abstract}
A model is presented
for generation of
fast solar wind in coronal holes,
relying on heating
that is dominated by
turbulent dissipation of MHD fluctuations
transported upwards
in the solar atmosphere.
Scale-separated transport equations
include
large-scale fields,
transverse Alfv\'enic fluctuations,
and a small compressive dissipation
due to parallel shears near the transition region.
The
model
accounts for proton temperature, density,
wind speed, and fluctuation amplitude
as observed in
remote sensing and \emph{in situ} satellite data.
\end{abstract}


        \section{Introduction}

An open question in solar and heliospheric physics
is to identify the physical processes responsible for heating the corona
and accelerating the fast solar wind streams emanating from coronal
holes.    This requires that a fraction
of the energy available in photospheric motions
be transported through the chromospheric
transition region, and dissipated in the corona.
The measured speeds of fast solar wind streams
require spatially extended
heating \citep{WithbroeNoyes77,HolzerLeer80,Withbroe88}.
The physical mechanisms for this
transport and dissipation
have remained elusive.
Some models have resorted
to use of a parametrically defined heat
deposition (a ``heat function'') that decays exponentially with height,
or anomalous heat conduction that
redistributes energy along field-lines
        \citep{HabbalEA95,McKenzieEA95,BanaszkiewiczEA98}.
One-dimensional (1D) models of this kind, extending
from the chromosphere to 1\,AU
        \citep{HansteenEA94,HansteenEA97},
have helped in
understanding the
regulation of the solar wind mass flux
and can reproduce fast solar wind
streams originating in cool electron coronal holes.

Here we present a
model that
demonstrates solar wind acceleration
due to heating by
a quasi-incompressible turbulent
cascade triggered by
coronal stratification \citep{MatthaeusEA99},
and supplemented by compressive heating
near the base. This model accounts
for most presently available coronal and
interplanetary observations.

The idea that broadband plasma fluctuations might 
heat the extended corona and accelerate the
solar wind has long been discussed
  \citep{Coleman68,BelcherDavis71,Hollweg86,HollwegJohnson88,Velli93a}.
However,
the mechanisms of transfer of fluctuation
energy to small scales---and, in particular,
the role of Alfv\'enic turbulence (as observed in the solar wind)
and cascade processes
  (e.g., phase mixing, ponderomotive driving, 
shocks, etc.)---have not
been described
self-consistently.
Two recent papers shed light on 
these relationships \citep{SuzukiInutsuka05,CranmerEA07}.

  \citet{SuzukiInutsuka05} incorporate
1D compressive nonlinear interactions driven by
Alfv\'en waves and leading to shock heating.
This model
produces good agreement with solar wind speed
profiles.
As low-frequency Alfv\'en waves propagate upward, their
wave pressure compresses the plasma.
Unable to refract or mode-mode couple into a perpendicular wavenumber
cascade, these waves must dissipate in 1D shock fronts.
This model provides a
valuable demonstration that MHD fluctuations can act as a conduit
to transport energy to the requisite altitudes.
However the restriction to 1D cascade is at odds with the
well-established propensity for an incompressible MHD cascade to
proceed mainly through wavevectors
perpendicular to a strong mean magnetic field
        \citep{RobinsonRusbridge71,ShebalinEA83,%
                OughtonEA94,BieberEA96,ChoEA02-strongB0}.
Furthermore the corona exhibits
a clear transverse structuring,
and the initial fluctuations must have perpendicular correlation
lengths not much larger than a super-granulation scale
        ($15\,000$\,km).

Another recent model
        \citep{CranmervanBall05,CranmerEA07}
incorporates a low-frequency cascade model
       \citep{VerdiniEA06-soho};
however, the
treatment of propagation and
dissipation differs
significantly from the present approach.
Their scheme
treats nonlinear effects as a perturbation,
and it is unclear if it converges for
strong turbulent heating.
Here we employ a strong turbulence closure.
We also do not rely
on electron heat conduction to boost radial
energy transport.
Instead we compute an internal energy associated
with the protons only.
This approach supports comparisons
with results employing
improved representation of turbulence,
such as shell models
        \citep{VerdiniEA09-emp, VerdiniEA09-apj}
and (potentially) full MHD simulation.

We find here that reflection of
Alfv\'enic turbulence alone
does not lead to a full corona/solar
wind stationary state --  a compressible
contribution is required.
This is supported by
recent observations
from \emph{Hinode}
        \citep{LangangenEA08,DePontieuEA09}.  
When we
include a small
component of compressive heating near the coronal base,
fast solar wind streams are then accounted for.
        \section{Model description and equations}
We employ a
        1D large-scale steady-state
MHD model
in an expanding flux tube
        \cite[cf.\ ][]{LeerEA82,VerdiniEA06-soho},
comprising
equations of mass continuity,
radial momentum conservation,
and pressure (internal energy),
\begin{eqnarray}
  \Deriv{}{r}
        \left[
           \rho U A
        \right]
        & = &
        0 ,
                                                    \label{eq:continuity} \\
  \rho U \Deriv{U}{r}
        & = &
        - \Grad P'
        - \rho \frac{GM_{\rm sun}}{r^2}
        + {\sR}_r ,
                                                    \label{eq:mtm} \\
  U \Deriv{p}{r}
        & = &
        - \gamma p \Div {\vU}
        + (\gamma - 1) Q(r) .
                                                    \label{eq:pressure}
\end{eqnarray}
Here $r$ is the radial coordinate,
 $ A(r)$ the flux tube cross-sectional area,
 $ p(r)=2 n k_{\text{B}}T$ the thermal pressure,
 $ \vU = U(r) \rhat $  the large-scale radial flow (wind) velocity,
 $ G$ the gravitational constant,
and
 $ M_{\textrm{sun}} $ the solar mass.
$\sR_r$ is the radial component of the (vector)
divergence of the
MHD Reynolds stress
  $ {\sbR} = \langle
        \delta \vb \cdot \Grad \delta \vb /4\pi
        -
        \rho \delta \vu \cdot \Grad \delta \vu
        \rangle
  $,
where
        $ \delta \vu $,
        $ \delta \vb $
are the fluctuations and the full magnetic field is
        $ \vB = \rhat B_r(r) + \delta \vb $, 
and $ \rhat \cdot \delta \vb = 0 = \rhat \cdot \delta \vu $.

The
total (thermal plus magnetic)
pressure is
\begin{equation}
   P' =
          2 n k_{\text{B}} T + \frac{\delta \vb^2}{2}.
   \label{eq:P-defn}
\end{equation}
We specify an area expansion factor $A(r)$.
Then $B_r(r)$ is
determined by magnetic flux
conservation,
        $ B_r(r) A(r) = const$,
which is constrained by 1 AU observations.

$ Q(r) $ is the heating per unit volume,
related to the heat function (per unit mass)
        $ H = Q/\rho $.
It involves an incompressible part
        $ H_i $,
associated with turbulence,
modeled here in
        K\'arm\'an--Taylor fashion
        \citep[e.g.,][]{KarmanHowarth38,MattEA04-Hc}.
There is also a
small compressive part
        $ H_c = Q_c/\rho $,
so  that
        $ H = H_i + H_c $.
Turbulence influences the large-scale flow through  $Q_i$,
wave pressure, and the Reynolds stress $\sR_r$,
and is modeled using
only a few free parameters.

The dominant
contributor to the turbulent heating,
the low-frequency quasi-incompressible
turbulence,
is evolved using a
transport equation and one-point closure
that depends upon the cross helicity.
The nonlinear phenomenological model
        \citep{DmitrukEA01-apj}
is involves the Els\"asser variables
${\bf z}^\pm = \delta \vu \mp \delta \vb /\sqrt{4 \pi \rho}$,
their associated energies
 $ E^\pm  = Z_\pm/4   = \langle |{\bf z}^\pm|^2 \rangle/4 $,
  ($\langle \cdots \rangle$ indicates an ensemble average)
and a common similarity (correlation) scale
        $ \lambda $.
The dimensionless cross helicity
        $ \sigma_c = (Z_+^2 - Z^2_-) / (Z^2_+ + Z^2_-) $
measures any excess inward or outward propagating-type
fluctuations.
The incompressible turbulent heating
model\citep{HossainEA95} is
\begin{equation}
   H_i (r)
        =
        \frac{Q_i(r)}{\rho}
        \equiv
        \frac{1}{2} \frac{Z_-Z_+^2 + Z_+Z_-^2}{\lambda}.
   \label{eq:diss}
\end{equation}
See   also
\citet{DobrowolnyEA80-prl},
        \citet{GrappinEA83},
and
        \citet{MattEA04-Hc}.

We include spatial transport
        \citep{VerdiniVelli07}
in a non-uniform wind with speed
        $ U(r) \rhat$.
The fluctuations are assumed to
be Reduced MHD-like
(i.e., perpendicular fluctuations, parallel gradients much weaker than
transverse ones) and represented by ``typical amplitudes'' of a given
frequency
        $z_\pm(\omega)$,
defined such that
        $ Z_\pm^2 = \int_\Omega \left( z_\pm^2/\omega \right) \d\omega $:
\begin{eqnarray}
    \left[ U \pm V_a \right]
        \Deriv{ z_\pm}{r} +  i \omega z_\pm
        & = &
       R_1^\pm z_\pm
     + R_2^\pm z_\mp
      - \frac{|Z_\mp|}{2\lambda}z_\pm.
                                                        \label{eq:genome}
\end{eqnarray}
The WKB [Eq.~(\ref{eq:R1})] and reflection [Eq.~(\ref{eq:R2})]
terms are related to
large-scale gradients by
\begin{eqnarray}
     R_1^\pm
        & = &
  -\half\left[U\mp V_a\right]
  \left(\Deriv{\log V_a}{r}+\Deriv{\log A}{r}\right)
            \label{eq:R1} , \\
     R_2^\pm
        & = &
   \half\left[U\mp V_a\right]
  \Deriv{\log V_a}{r} .
         \label{eq:R2}
\end{eqnarray}
Here
        $V_a = V_{a_r} = B_r / \sqrt{4\pi\rho} $.
For simplicity the lengthscale $\lambda$ is associated
with the expansion, using
        $ \lambda (r) = \lambda_\odot  \sqrt{A(r)} $.
The
        $\odot$
subscript indicates evaluation at the coronal base, here taken
to be at the top of the transition region.

The smaller, compressive contribution
to the heating ($ H_c$)
is assumed to
be confined
near the coronal base
where it
rapidly dissipates through shocks.
This is motivated
by recent observations
        \citep{LangangenEA08,DePontieuEA09}  
of fluctuations with parallel (vertical) variance that
pervade the entire corona near the
transition region.
Here we model this
effect directly
as a heat function, and
assume that it contributes
 $ \approx 1\% $
of the total (height-integrated)
heat function.

Having in hand the
complete set of equations
(\ref{eq:continuity})--(\ref{eq:pressure}), and
(\ref{eq:genome}), along with
constitutive relations
(\ref{eq:P-defn}), (\ref{eq:diss}), (\ref{eq:R1}), and (\ref{eq:R2}),
and the small
compressive heat function $H_c$,
we are in position to
solve a relatively complete and
self-consistent solar wind model with turbulence.
The stationary solutions for the large-scale fields and turbulence
parameters are obtained
via a numerical iteration procedure in which
        $ T_\odot = 4 \times10^{5} \, \mathrm{K}$
and
        $ n_\odot = 5 \times10^{8} \, \mathrm{cm^{-3}} $
are held constant.
To begin,
we impose trial radial profiles for $T$ and $n$,
and then solve
        Eqs.~(\ref{eq:genome})
with a fixed amplitude of the velocity fluctuation at the base, say
        $ \delta u_\odot = 30 \, \mathrm{km \, s^{-1}} $.
From this temporary solution we compute the
Reynolds stress, the ponderomotive force, and the heating function
that appear in
        Eqs.~(\ref{eq:continuity})--(\ref{eq:pressure}).
Next, those equations are solved for
        $T$ and $U$,
which provides updated values for
        $ V_a $,
        $ U $,
and their derivatives.
These updated profiles are used in re-solving
        Eqs.~(\ref{eq:genome}),
and so on.

After few
iterations, there
are small relative differences
        ($ \lesssim 10^{-5}$)
in each of the large-scale
and fluctuation
fields.
We control convergence
through the relative variation of the positions of the
sonic and Alfv\'enic critical points, assuming convergence when
        $ \Delta r_s/r_s $,
        $ \Delta r_a/r_a < 10^{-4} $.
For each frequency
        $ \omega $,
we solve subject to imposed conditions on
        $ \delta u_\odot $,
        $ H_{c}(r) $,
the maximal (over)expansion
        $ \fmax = \max[A(r) / r^2] $,
and the correlation length scale
 	$ \lambda_\odot $.
We use, as in
        \citet{KoppHolzer76},
        $ f(r) =A(r)/r^2 $, with
\begin{equation}
 f(r) = \frac{\fmax\exp[(r-r_f)/\sigma] + f_1}
             {\exp[(r-r_f)/\sigma]      + f_1},
\end{equation}
$ f_1 = 1 - \fmax \exp[(\Rsun-r_f)/\sigma] $,
keeping fixed the location and
width of the super-radial expansion,
        $r_f = 1.31 \, \Rsun$
and
        $\sigma = 0.51 \, \Rsun$ respectively.
        \section{Results}
\begin{figure*}
 \begin{centering}
   \includegraphics[width=0.95\textwidth]{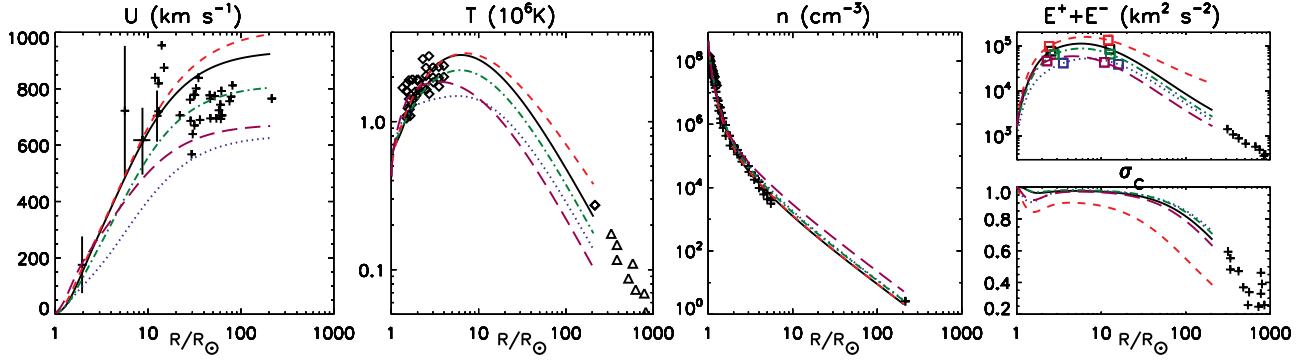}
  \caption{Wind speed, temperature, density, turbulence level,
          and normalized cross helicity as a function of distance
          for the representative solution (solid line) with:
                $\delta u_\odot = 30    \, \mathrm{km \, s^{-1}} $,
                $        \omega =  0    \, \mathrm{Hz} $,
                $ \lambda_\odot = 0.015 \, \Rsun $,
                $ f_{\mathrm{max}} = 12.5$,
        and
                $H_{c} = 1\% H $.
          The other lines denote solutions for
         which one of the parameters has been varied; specifically,
                $\delta u_\odot = 20   \, \mathrm{km \, s^{-1}} $,
                $ \lambda_\odot = 0.05 \, \Rsun $,
                $ f_{\mathrm{max}} = 10$,
        and
                $ H_c \approx 5\% H $,
        in blue dotted, red dashed, green dashed-dotted, violet long-dashed
        lines respectively.
        Empty squares in the top-right panel mark the position of
	$r_s$ and $r_a$ ($r_s<r_a$).
	The other symbols represent observational constraints for
        the fast solar wind taken from:
        \citet{McComasEA00}     for $U$, $n$ at 1\,AU;
        \citet{GrallEA96}       for $U$ inside 1\,AU;
        \citet{BreechEA05}      for $\sigma_c$
                and \emph{proton} temperature beyond 1\,AU;
        \citet{Cranmer09}       for \emph{proton} temperature inside 1\,AU;
        \citet{BanerjeeEA98} and \citet{FisherGuhathakurta95}
                                for $n$ inside 1\,AU;
        \citet{BavassanoEA00b}  for the turbulence level.
        }
  \label{fig:1}
 \end{centering}
\end{figure*}

\begin{figure}
\begin{centering}
  \includegraphics[width=.9\columnwidth] {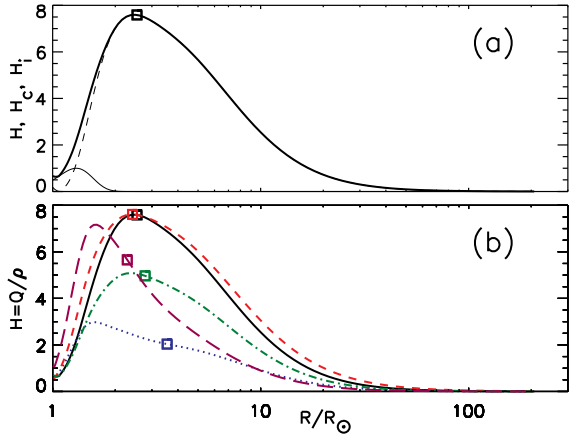}
    \caption{\emph{(a)}.
          Total (thick line), compressive (thin line) and incompressible
          (dashed line) heating function
          (dissipated energy per unit mass) as a
          function of heliocentric distance in
          units of
             $ \approx 3 \times 10^{10} \, \mathrm{cm^{2} \, s^{-3}} $
          for the reference solution.
          \emph{(b)}. Total heating function for
          the solutions obtained varying the parameters as described in the
          caption of Fig.~\ref{fig:1} (same line and color coding).
          Empty squares mark the location of $r_s$ for each solution.}
  \label{fig:2}
\end{centering}
\end{figure}

A typical solution is shown in
        Figure~\ref{fig:1} (solid black line),
where we plot several quantities
        ($U$, $T$, $n$, $\sigma_c$, $Z^2$)
as a function of distance, obtained
with the chosen parameters:
        $ \delta u_\odot = 30 \, \mathrm{km \, s^{-1}}$,
        $         \omega = 0  \, \mathrm{Hz}$,
        $  \lambda_\odot = 0.015 \, \Rsun $,
        $ f_{\mathrm{max}} = 12.5$,
and
        $ H_{c} = 1\%H $.
The speed $U$,
temperature $T$,
and density $n$
are generally in good agreement with the observations in the
heliosphere,
although the terminal speed is about
        $100 \, \mathrm{km \, s^{-1}}$ higher
and
$T$ peaks too far
from the coronal base compared to the observed \emph{proton} temperature.
The latter is obtained by subtracting from the observed line width
$w_\bot$ the contribution of the turbulent fluctuation $\delta u$,
according to the relation\footnote{%
    Only the reference case is used to
    compute the corrected proton temperature in Fig.~\ref{fig:1}. 
    Other cases give similar values except the
        $ \delta u_\odot = 20 \, \mathrm{km \, s^{-1}} $ and
        $ H_c = 5\%H $ cases,
    yielding much higher corrected temperatures. 
         }
$2k_BT/m_p = w_\bot^2 - (\delta u)^2/2$
        \citep[e.g.,][]{TuEA98}.
The profiles of the
fluctuation energy
        $ E^+ + E^- = Z^2/2 $
and normalized cross helicity
        $ \sigma_c $
trend nicely towards the \emph{in situ} data beyond
        1\,AU.
The turbulent dissipation $H_i$ (Fig.~\ref{fig:2}a, dashed line)
accounts for
the spatially extended heating that
accelerates the wind from below the
sonic critical point.
The compressive heating is small,
about
        1\%
of the total heating per unit mass.
Its form is arbitrarily chosen to be a gaussian
        (Fig.~\ref{fig:2}a, thin solid line),
centered at
        $ 1.3  \, \Rsun $
with width
        $ \approx 0.25 \, \Rsun $,
in order to confine its contribution well below $r_s$.

The solutions change as parameters are varied,
but not all the parameters have the same impact.
For example, for
a steep spectrum of low-frequency waves
  ($ \omega \lesssim 10^{-5} \, \mathrm{Hz}$,
         slope~$\lesssim -1.1$),
use of only zero-frequency
fluctuations is a very good approximation.
For flatter spectra, the high-frequency
part of the spectrum
remains principally
outward propagating [$ z^+(\omega) $],
thus limiting the total turbulent dissipation.
This affects the turbulent energy
        $ Z^2 $,
which becomes larger beyond
        $ r_a \approx 13 \, \Rsun $,
and also
        $\sigma_c$,
which stays closer to unity.
$U$, $T$, and $\rho$
are almost unchanged ($U$ and $T$ are slightly reduced).
Below, we consider only $\omega=0$
fluctuations.

The position of the sonic critical point
depends upon
momentum and heat
addition
  \citep{LeerEA82}.
Given that we specify $A_\odot$ and $\rho_\odot$, the mass
flux is determined by $U_\odot$,
which is found by requiring that the solution
becomes supersonic on passing through $r_s$.
It follows that deposition of heat
before $r_s$ increases
        $ T( r_s ) $
and
typically $U_\odot$, and thus also the mass flux.
Deposition of heat beyond $r_s$ does not alter
        $ T( r_s ) $,
and so the mass flux is unchanged;
however $U$ increases and $\rho$ decreases beyond $r_s$,
with respect to a reference solution.
Generally speaking, the maximum of
        $ Z^2 $
is below
        $r_a$
(it coincides with $r_a$ in the undamped case),
the turbulent heating peaks near $r_s$,
and the momentum added by the wave
pressure
has a maximum inside $r_s$.

Two ways of controlling the heat and momentum deposition
[i.e.,
adjusting the importance of nonlinear terms relative
to reflection/WKB (linear) terms in
        Eqs.~(\ref{eq:genome})],
are through variations of
        $ \delta u_\odot$ and $\lambda(r)$.
Decreasing the basal fluctuation amplitude
        ($\delta u_\odot = 20 \, \mathrm{km \, s^{-1}}$,
        in
        Fig.~\ref{fig:1})
decreases the turbulent heating and acceleration.
In addition heat deposition
peaks well below $r_s$ (Fig.~\ref{fig:2}b), yielding a slow,
overdense, cool wind, with a deficit of turbulent
energy. It is clear that $H_c$ shapes the solution only below
        $ 1.5 \, R_\odot$
while differences arise from a different shape of $H_i$.
The asymptotic
        $ Z^2 $
and
        $ \sigma_c $
do not vary much since beyond $r_a$ reflection is negligible and
the ratio $Z_-/Z_+$ is then controlled by
turbulent dissipation and $\lambda(r)$.

Modification of $\lambda(r)$
leaves unchanged the position of the
critical points;
i.e, it only slightly alters the turbulent heating
close to the coronal base, where reflection controls the ratio $Z_-/Z_+$.
A larger basal correlation scale
        ($\lambda_\odot = 0.05 \, \Rsun$
        in Fig.~\ref{fig:1})
affects the solution mainly \emph{beyond} $r_s$,
yielding a hotter and faster wind, but with the density almost unchanged.
The fluctuations have features similar to the undamped solutions,
resulting in excess turbulent energy (that peaks closer to
$r_a$) and more inward propagating waves (a smaller $\sigma_c$).
Nonetheless, the increased turbulent energy---driven
        by reflection and the WKB term---broadens
the turbulent heating
while keeping fixed its maximum
        (Fig.~\ref{fig:2}b).

Adjusting the expansion $A(r)$
also
changes the correlation scale
through $ \lambda(r) = \lambda_\odot \sqrt{A(r)} $.
$A(r)$
controls the reflection term directly and
through the Alfv\'en speed
gradients
        [Eqns.~(\ref{eq:R1})--(\ref{eq:R2})].
Decreasing the maximum super-radial expansion,
        to $ f_{\mathrm{max}} = 10 $,
while
keeping fixed its location at
        $ r_f \approx 1.3 \, \Rsun < r_s $
        (Fig.~\ref{fig:1})
has two consequences.
First, it decreases the correlation scale for
        $ r \gtrsim r_f$,
causing a reduction of turbulent energy and heating.
Second, it increases the density scale height around
        $r_f$,
hence reducing reflection and the amount of
turbulent dissipation there.
The result is much smaller turbulent heating that peaks again at $r_s$
(Fig.~\ref{fig:2}b) yielding
a slower, cooler, and denser wind,
even as close as
	$r_s$,
which retains good asymptotic
	$Z^2$
and
	$\sigma_c$.

Finally, let us examine the role of 
compressive heating,
recalling that
        $ H(r) = H_c + H_i $.
If the compressive heating is increased to
        $ H_c \approx 5\%H $
        (Fig.~\ref{fig:1}, violet dashed),
the temperature maximum
moves closer to the coronal base, 
but the resulting wind is slower,
denser, and cooler.
An increased $H_c$ also alters
the
incompressible heating---the height-integrated heating is almost
unchanged, but heat deposition
occurs at lower $r$ due to increased reflection at $r< r_s$
        (Fig.~\ref{fig:2}b,
compare the smaller $\sigma_c$ in Fig.~\ref{fig:1}).
Beyond the peak of $H$ the fluctuation energy remains small, 
reducing the extended turbulent heating.

On the other hand, decreasing
        $ H_c $
enhances the incompressible heating; then
$H$ has a minimum near the maximum of 
$V_a$, just inside
$r_s$, and also peaks outside $r_s$. 
This causes convergence problems.
In early iterations, heat deposition is mainly outside $r_s$,
yielding a very fast, underdense, and hot wind,
with strong density gradient (and hence reflection) at $r> r_s$.
This leads to runaway iterations in which the sonic
point moves to larger $r$ with a temperature minimum
inside $r_s$.
When the temperature minimum becomes 
very low the sonic critical point
is reached with $\d T/\d r <0$.
Then steep gradients occur close
to the coronal base, heat deposition occurs close
to $r_s$, and the solution again resembles
the initial solution, producing the ensuing runaway.
To obtain a solution, 
the key features of $H_c$ are that it is
localized well below $r_s$,  and 
is at least $\sim 0.7 \% $ of the total heating
(for other parameters at reasonable levels). 
We found that the results do not depend strongly on the form of the compressive
heat function provided that this heating is localized and not too large
(exponential, gaussian, etc. all work).
        \section{Discussion}
The above model shows that turbulence near the coronal base,
originating through chromospheric transmission of fluctuations,
can heat the plasma in an expanding coronal hole flux-tube
and produce a fast solar wind that matches a number of observational
constraints.
The turbulence is mainly of the low-frequency Alfv\'enic
type.  A small amount of compressive heating between
the transition
region and the sonic point
appears to be needed to match the observations.
This additional heating may be due to type II spicules that supply
broadband low-frequency vertical fluctuations at
transition region heights, thus launching compressive
MHD modes near the coronal base
        \citep{DePontieuEA09}.

Most of the
fluctuation energy
is in low-frequency turbulence,
and this sustains
a strong anisotropic MHD cascade
through reflections from local density gradients.
This type of anisotropic cascade
is favored in MHD turbulence in the
presence of a strong DC magnetic field
        \citep{RobinsonRusbridge71,ShebalinEA83,OughtonEA94}.
Heat conduction does not enter the present
model at all, since it mainly affects electron
internal energy,
which evolves
independently in this approximation.
Similar assumptions work well in understanding
  observations of solar wind turbulence
        \citep{BreechEA09,CranmerEA09}.
In these ways, the present model differs
from other recent models that
incorporate turbulent heating
       \citep{SuzukiInutsuka05,CranmerEA07}.
In particular we believe that this
model demonstrates, possibly for the first time, that
a model
(almost) free of ad hoc
heat functions, artificial equations of state,
and
ad hoc assumptions about heat conduction,
can indeed heat the corona and accelerate the solar wind.

We plan further
study of the required small amount of
compressible heating, attributed here to
spicule-driven magnetosonic activity.
Another	useful extension
would be
to include separate
electron and proton internal energy budgets,
which will enable
additional observational constraints, and
will permit study of the role of kinetic dissipation processes
        \citep{BreechEA09,CranmerEA09}.
\acknowledgments
This research was supported in part by
NASA (Heliophysics Theory, NNX08AI47G) and
NSF (Solar-Terrestrial and SHINE programs ATM-0539995 and
ATM-0752135). AV acknowledges support from the Belgian Federal Science Policy
Office through the ESA-PRODEX program. 
We thank the UVCS project (J. Kohl, S. Cranmer) for providing data used in
Fig.~\ref{fig:1}.


\end{document}